\documentclass[aps,prd,reprint,twocolumn,superscriptaddress,longbibliography,nofootinbib,floatfix,showpacs]{revtex4-1}
\usepackage{epsfig}
\usepackage{amsmath,amssymb,amsfonts}
\usepackage{hyperref}
\usepackage{mathrsfs}
\usepackage{bbm}
\usepackage{slashed}
\usepackage{graphicx}
\usepackage{verbatim}

\usepackage{bm}

\usepackage[usenames]{color}

\usepackage{tikz}
\usetikzlibrary{calc,decorations.markings,decorations.pathmorphing,positioning,shapes,snakes,arrows}

\definecolor{darkgreen}{rgb}{0.2,0.6,0}

\newcommand{\be}{\begin{equation}}
\newcommand{\ee}{\end{equation}}
\newcommand{\bw}{\begin{widetext}}
\newcommand{\ew}{\end{widetext}}
\newcommand{\bi}{\begin{itemize}}
\newcommand{\ei}{\end{itemize}}
\newcommand{\bea}{\begin{eqnarray}}
\newcommand{\eea}{\end{eqnarray}}
\newcommand{\ud}{\mathrm{d}}

\newcommand{\LCm}{{\scriptscriptstyle -}} 
\newcommand{\LCp}{{\scriptscriptstyle +}}
\newcommand{\LCpm}{{\scriptscriptstyle \pm}}
\newcommand{\LCmp}{{\scriptscriptstyle \mp}}

\newcommand{\LCperp}{{\scriptscriptstyle \perp}}

\newcommand{\LCpara}{{\scriptscriptstyle \parallel}}

\usepackage[T1]{fontenc} \usepackage[latin1]{inputenc}

\begin{document}

\title{Resummation of quantum radiation reaction in plane waves}

\author{Greger Torgrimsson}
\email{g.torgrimsson@hzdr.de}
\affiliation{Helmholtz-Zentrum Dresden-Rossendorf, Bautzner Landstra{\ss}e 400, 01328 Dresden, Germany}

\begin{abstract}

We propose a new approach to obtain the momentum expectation value of an electron in a high-intensity laser, including multiple photon emissions and loops. We find a recursive formula that allows us to obtain the $\mathcal{O}(\alpha^n)$ term from $\mathcal{O}(\alpha^{n-1})$, which can also be expressed as an integro-differential equation. In the classical limit we obtain the solution to the Landau-Lifshitz equation to all orders. We show how spin-dependent quantum radiation reaction can be obtained by resumming both the energy expansion as well as the $\alpha$ expansion. 
	
\end{abstract}	

\maketitle


An electron in an electromagnetic field emits photons and the recoil it experiences is called radiation reaction (RR)~\cite{Burton:2014wsa,Blackburn:2019rfv}. Classically, the standard equation is the Abraham-Lorentz-Dirac (LAD) equation, but since it leads to unphysical solutions it is common to replace it with the Landau-Lifshitz (LL) equation. The first RR experiments with high-intensity lasers were performed recently~\cite{Cole:2017zca,Poder:2018ifi} (see also~\cite{Wistisen:2017pgr}), and in upcoming experiments one may expect significant quantum effects~\cite{Blackburn:2015tva,VranicQuantumRR,Dinu:2015aci,Neitz:2013qba,Yoffe:2015mba}. In this paper we propose a new method for obtaining quantum RR in high-intensity lasers. It is based on incoherent products of $\mathcal{O}(\alpha)$ terms (cf.~\cite{DiPiazza:2010mv}), with Mueller matrices taking spin into account (cf.~\cite{Seipt:2018adi,Geng:2020hcx}).

Let $p_\mu$ and $P_\mu$ be the initial and final momentum of the electron, $k_\mu$ the wave vector of the plane wave, which is described by a gauge potential $a_\LCperp(\sigma)$ where $\sigma=kx=\omega(t+z)$ is the (rescaled) lightfront time and $a_\LCpm=0$\footnote{We use units with $m_e=1$, $E$ is the field strength and $\omega$ a characteristic frequency scale, we absorb the electron charge into the background field ($eE\to E$), and lightfront coordinates are defined by $v^\LCpm=2v_\LCmp=v^0\pm v^3$ and $v_\LCperp=\{v_1,v_2\}$}. We will focus on $\langle kP\rangle=\sum_{n=0}^\infty\langle kP\rangle^{(n)}$, where $\langle kP\rangle^{(n)}$ is $\mathcal{O}(\alpha^n)$ and the dependence on the initial and final spin can be expressed as $\langle kP\rangle^{(n)}=(1/2){\bf N}_0\cdot{\bf M}^{(n)}\cdot{\bf N}_f$, where ${\bf N}=\{1,{\bf n}\}$ is the Stokes vector for spin along the unit vector ${\bf n}$ and ${\bf M}^{(n)}$ are ``strong-field-QED Mueller matrices''. Averaging and summing over the initial and final spin gives $\{1,{\bf 0}\}\cdot{\bf M}\cdot\{1,{\bf 0}\}$. 
We know from~\cite{Dinu:2018efz,Dinu:2019pau,Torgrimsson:2020gws} how to obtain $\mathcal{O}(\alpha^n)$ probabilities to leading order for long pulses or large $a_0=E/\omega$ by multiplying $\mathcal{O}(\alpha)$ Mueller matrices. In this case we need ${\bf M}^{\rm C}$ for (nonlinear) Compton scattering and ${\bf M}^{\rm L}$ for the cross term between the $\mathcal{O}(\alpha^0)$ and $\mathcal{O}(\alpha)$ parts of the amplitude for $e^\LCm\to e^\LCm$.
From this we now find the following recursive formula
\be\label{generalRecursive}
\begin{split}
{\bf M}^{(n)}(b_0,\sigma)=&\int_\sigma^\infty\!\frac{\ud\sigma'}{b_0}\!\int_0^1\!\!\ud q({\bf M}^{\rm L}(b_0,\sigma',q)\cdot{\bf M}^{(n-1)}(b_0,\sigma') \\
&+{\bf M}^{\rm C}(b_0,\sigma',q)\cdot{\bf M}^{(n-1)}(b_0(1-q),\sigma')) \;,
\end{split}
\ee 
where $b_0=kp$, $q=kl/kp$ and $l$ is the photon momentum, and ${\bf M}^{(0)}=b_0{\bf 1}$ (${\bf M}^{(0)}={\bf 1}$ for the probability). The lower integration limit has been introduced so that the products of Mueller matrices are lightfront-time ordered. The final result is obtained by setting $\sigma=-\infty$. The shift $b_0\to b_0(1-q_1)$ in the Compton term takes RR into account.

With ${\bf M}=\sum_{n=0}^\infty{\bf M}^{(n)}$, \eqref{generalRecursive} implies
\be\label{integro-diff-eq}
\frac{\partial{\bf M}}{\partial\sigma}=-\int_0^1\frac{\ud q}{b_0}({\bf M}^{\rm L}\cdot{\bf M}(b_0)+{\bf M}^{\rm C}\cdot{\bf M}(b_0[1-q])) \;.
\ee
This integro-differential matrix equation gives quantum RR to all orders in $\alpha$ with spin taken into account. Note that even if we are only interested in unpolarized initial and final states, we still need to solve a matrix equation before we can project with $\{1,{\bf 0}\}\cdot{\bf M}\cdot\{1,{\bf 0}\}$. Note also that~\eqref{integro-diff-eq} holds even if $a_0$ is not large, provided the field is sufficiently long and the full version of ${\bf M}^{\rm C,L}$ is used.
Eq.~\eqref{integro-diff-eq} looks similar to kinetic RR equations~\cite{Neitz:2013qba} (see also~\cite{cascadeEqs}), so one may expect that it can be solved with similar methods.
However, in this paper we will instead use~\eqref{generalRecursive} to obtain the first ${\bf M}^{(n)}$ terms before performing a resummation.


We will first show that the classical limit of~\eqref{generalRecursive} agrees with the solution to LL. Classically there is no spin dependence, so we neglect the spin components and focus on $M=\{1,{\bf0}\}\cdot{\bf M}\cdot\{1,{\bf0}\}$. Using $M^{\rm L}=-M^{\rm C}$ and expanding to first order in the photon momentum, we find
\be
\begin{split}
M^{(n)}(b_0,\sigma)&\approx\int_{\sigma}^\infty\ud\sigma'\int_0^1\!\ud q\, M^{\rm C}(-q)\frac{\partial M^{(n-1)}}{\partial b_0} \\
&=-\int_{\sigma}^\infty\ud\sigma'\frac{\partial M^{(1)}}{\partial\sigma'}\frac{\partial M^{(n-1)}}{\partial b_0}(b_0,\sigma') \;,
\end{split}
\ee
which is solved by
\be
\begin{split}
M^{(n)}(b_0,\sigma)=&n! b_0^{1+n}\int_\sigma^\infty\ud\sigma_1\int_{\sigma_1}^\infty\ud\sigma_2...\int_{\sigma_{n-1}}^\infty\ud\sigma_n \\
\times&J(\sigma_1)...J(\sigma_n)=b_0\left(b_0\int_\sigma^\infty\!\ud\sigma' J(\sigma')\right)^n \;,
\end{split}
\ee
where $J$ is obtained by matching with $M^{(1)}$,
which we have already calculated in~\cite{Ilderton:2013tb,Ilderton:2013dba},
\be
\langle kP\rangle^{(1)}\approx-b_0\Delta\int\ud\sigma\,{\bf a}'^2 \qquad \Delta=\frac{2}{3}\alpha b_0 \;.
\ee
Resumming this geometric series we finally find
\be
\langle kP\rangle=\sum_{n=0}^\infty\langle kP\rangle^{(n)}=\frac{b_0}{1+\Delta\int\ud\sigma\,{\bf a}'^2} \;.
\ee 
In~\ref{ClassicalLimitPerpPlus} we show that the $\langle P_{\LCperp,\LCp}\rangle$ components can be calculated in a similar way. We find
\be\label{mperpLL}
\langle P_{\LCm,\LCperp}\rangle=\pi_{\LCm,\LCperp}+\frac{\Delta}{1+\Delta\int\ud\sigma\,{\bf a}'^2}\left[\pi'-\int\ud\sigma\,{\bf a}'^2\pi\right]_{\LCm,\LCperp}
\ee
and $\langle P_\LCp\rangle$ satisfies the mass-shell condition $P_\LCp=(P_\LCperp^2+1)/(4P_\LCm)$.
This resummation\footnote{The fact that an $\alpha$ expansion would have to be resummed to obtain LL has recently been considered in~\cite{Heinzl:2021mji}.} agrees exactly with the exact solution to LL~\cite{DiPiazzaLLsol} (see also~\cite{Neitz:2013qba}).

This does not mean that LAD does not agree with the classical limit of QED, because our approximation neglects terms that have subdominant scaling with respect to the pulse length and/or $a_0$. From Eq.~(4.39) in~\cite{Ilderton:2013dba} we see that the difference between LL and LAD at $\mathcal{O}(\alpha^2)$ is a term proportional to ${\bf a}'^2$, i.e. a term without an integral. This term vanishes at asymptotic $\sigma$, but is also subdominant at finite times because it has no pulse-length scaling. We can therefore not rule out that the classical limit of all QED contributions may agree with LAD rather than LL. 

In the limit of a very long pulse we have
\be\label{largeTclassical}
\langle P_\LCm\rangle\approx\frac{p_\LCm}{\Delta\int{\bf a}'^2} \qquad \langle P_\LCperp\rangle\approx\frac{\int{\bf a}'^2(a-\hat{a})_\LCperp}{\int{\bf a}'^2} \;,
\ee
so $\langle P_\LCm\rangle$ becomes small, $\langle P_\LCperp\rangle$ stays $\mathcal{O}(1)$ and $\langle P_\LCp\rangle$ becomes large, and (since $\Delta\propto p_\LCm$) all components of $\langle P_\mu\rangle$ become independent of the initial momentum.


\begin{figure}
\includegraphics[width=\linewidth]{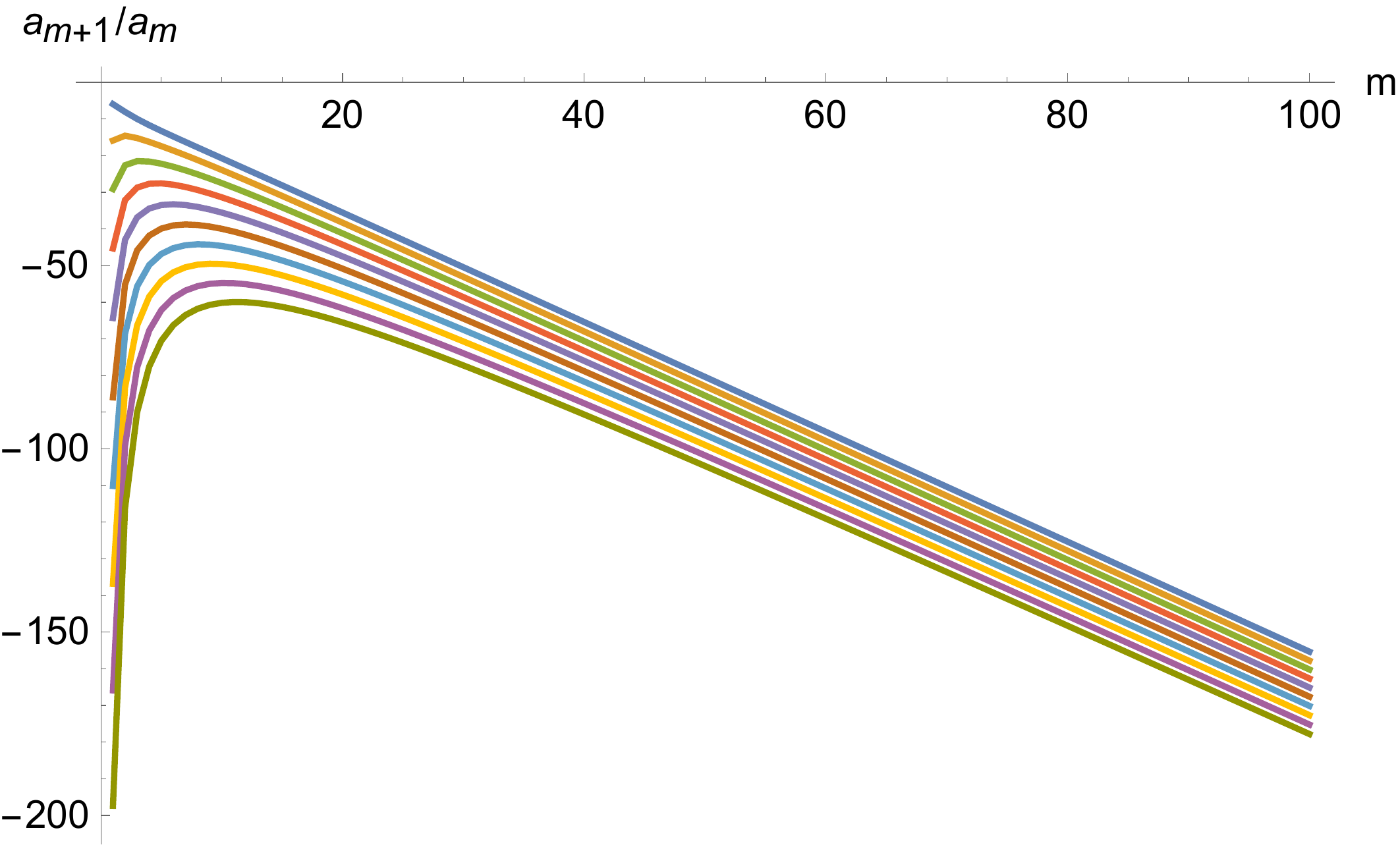}
\caption{The ratios of neighboring coefficients in the $\chi$ expansion for $\{1,0\}\cdot{\bf a}_m$. The different lines correspond to $\langle kP\rangle^{(1)}$ (top) to $\langle kP\rangle^{(10)}$ (bottom). The linear scaling at higher orders shows that the coefficients grow factorially and the $\chi$ expansion is hence asymptotic.}
\label{coefRatioFig}
\end{figure}

\begin{figure}
\includegraphics[width=\linewidth]{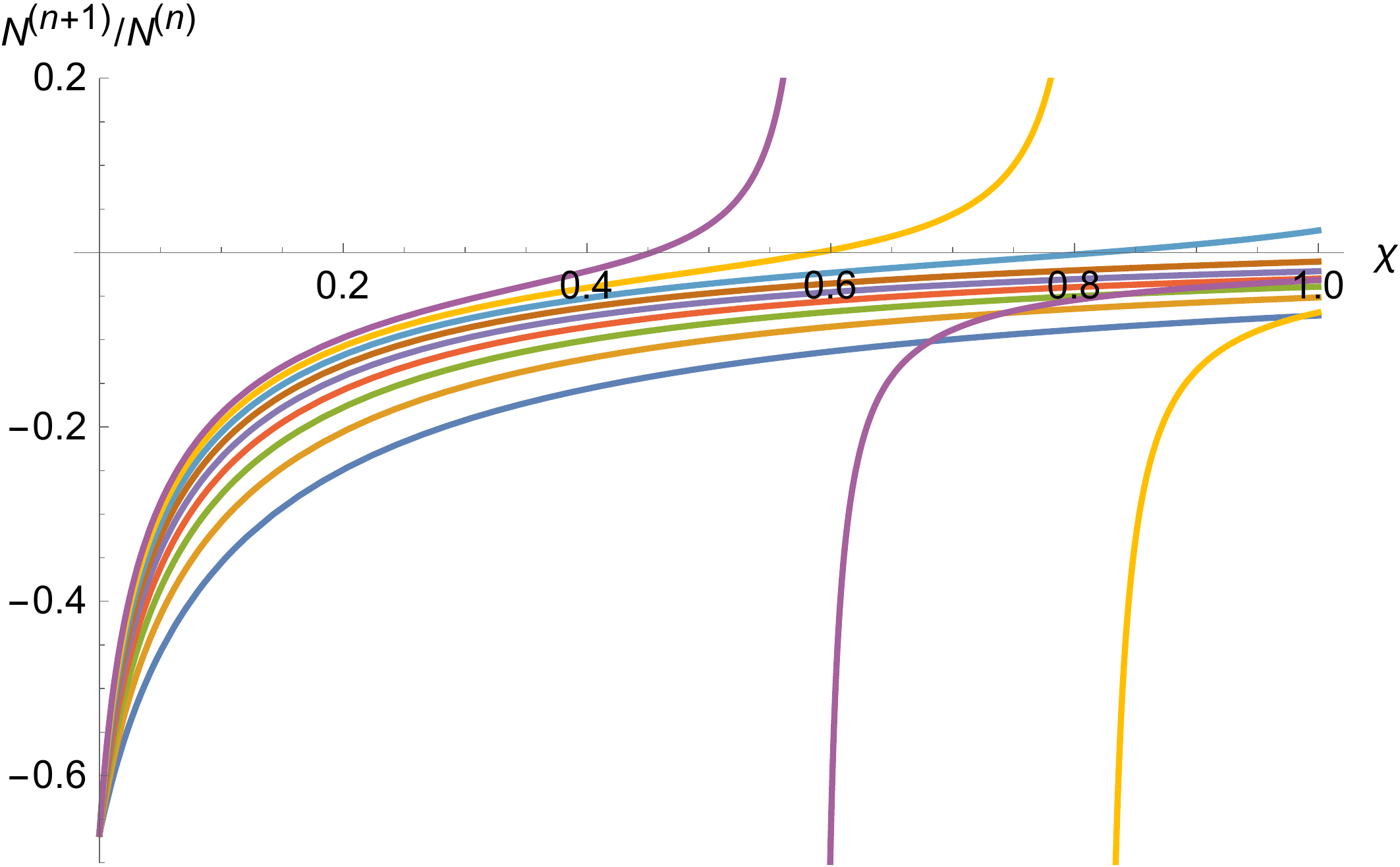}
\caption{The ratios $N^{(n+1)}/(\chi N^{(n)})$ of neighboring coefficients in the $\alpha$ expansion as a function of $\chi$, for an unpolarized initial state, $N^{(n)}=\{1,0\}\cdot{\bf N}^{(n)}$. The ratios have been divided by $\chi$ in order to show the convergence to the classical limit $N^{(n+1)}/N^{(n)}\to-2\chi/3$ for $\chi\to0$.}
\label{alphaRatiosFig}
\end{figure}

Having shown that~\eqref{generalRecursive} gives the expected classical limit, we now turn to quantum RR. For simplicity we focus on $\langle P_\LCm\rangle$, and we sum over the final spin, i.e. ${\bf N}_f=\{1,{\bf 0}\}$ is fixed and we have an overall factor of $2$, so we use ${\bf N}^{(n)}:={\bf M}^{(n)}\cdot\{1,{\bf 0}\}$.
In this first application we consider a constant field, so the $\sigma$ integrals give $\int\ud\sigma_n...\ud\sigma_1\to(\Delta\phi)^n/n!$, and it is convenient to rescale ${\bf N}^{(n)}$ so that
\be
a_0\langle kP\rangle^{(n)}=T^n{\bf N}_0\cdot{\bf N}^{(n)}
\ee
where $T=\alpha a_0\Delta\phi$ is our effective expansion parameter, which can be $T>1$ for large $a_0\Delta\phi$. 
In general ${\bf N}$ has four elements, but here we consider only initial and final states that are either unpolarized or polarized (anti-)parallel to the magnetic field, and then only two elements contribute. Omitting the irrelevant elements we have ${\bf N}_0=\{1,0\}$ and ${\bf N}_0=\{1,\pm1\}$ for unpolarized and polarized states. We have
\be\label{recursionLCF}
{\bf N}^{(n)}=\int_0^1\frac{\ud q}{n\chi}({\bf M}^{\rm C}\cdot{\bf N}^{(n-1)}(\chi(1-q))+{\bf M}^{\rm L}\cdot{\bf N}^{(n-1)}(\chi)) \;,
\ee
where ${\bf N}^{(0)}(\chi)=\chi\{1,0\}$, $1/n$ comes from $(\Delta\phi)^n/n!$, $\chi=a_0b_0$ is the quantum nonlinearity parameter (gives the electric field in the rest frame of the electron), 
\be
{\bf M}^{\rm C}=\begin{pmatrix}-\text{Ai}_1(\xi)-\kappa\frac{\text{Ai}'(\xi)}{\xi} & \frac{q}{s_1}\frac{\text{Ai}(\xi)}{\sqrt{\xi}} \\ q\frac{\text{Ai}(\xi)}{\sqrt{\xi}} & -\text{Ai}_1(\xi)-2\frac{\text{Ai}'(\xi)}{\xi}\end{pmatrix}
\ee
and
\be
{\bf M}^{\rm L}=\begin{pmatrix}\text{Ai}_1(\xi)+\kappa\frac{\text{Ai}'(\xi)}{\xi} & -q\frac{\text{Ai}(\xi)}{\sqrt{\xi}} \\ -q\frac{\text{Ai}(\xi)}{\sqrt{\xi}} & \text{Ai}_1(\xi)+\kappa\frac{\text{Ai}'(\xi)}{\xi}\end{pmatrix} \;,
\ee
where $\xi=(r/\chi)^{2/3}$ with $r=(1/s_1)-1$, $\kappa=(1/s_1)+s_1$ and $s_1=1-q$.  

In order to compute ${\bf N}^{(n)}$ we have used two completely different methods. In the first we compute ${\bf N}^{(1)}(\chi)$ between $\chi=0$ and some\footnote{The probability that one of the emitted photons decays into a pair (trident pair production) scales as $e^{-16/(3\chi)}$ \cite{Ritus:1972nf}, which should be small.} $\chi_{\rm max}$ and make an interpolation function of it, which is then used in~\eqref{recursionLCF} to compute an interpolation function of ${\bf N}^{(2)}(\chi)$, and so on.


\begin{figure}
\includegraphics[width=\linewidth]{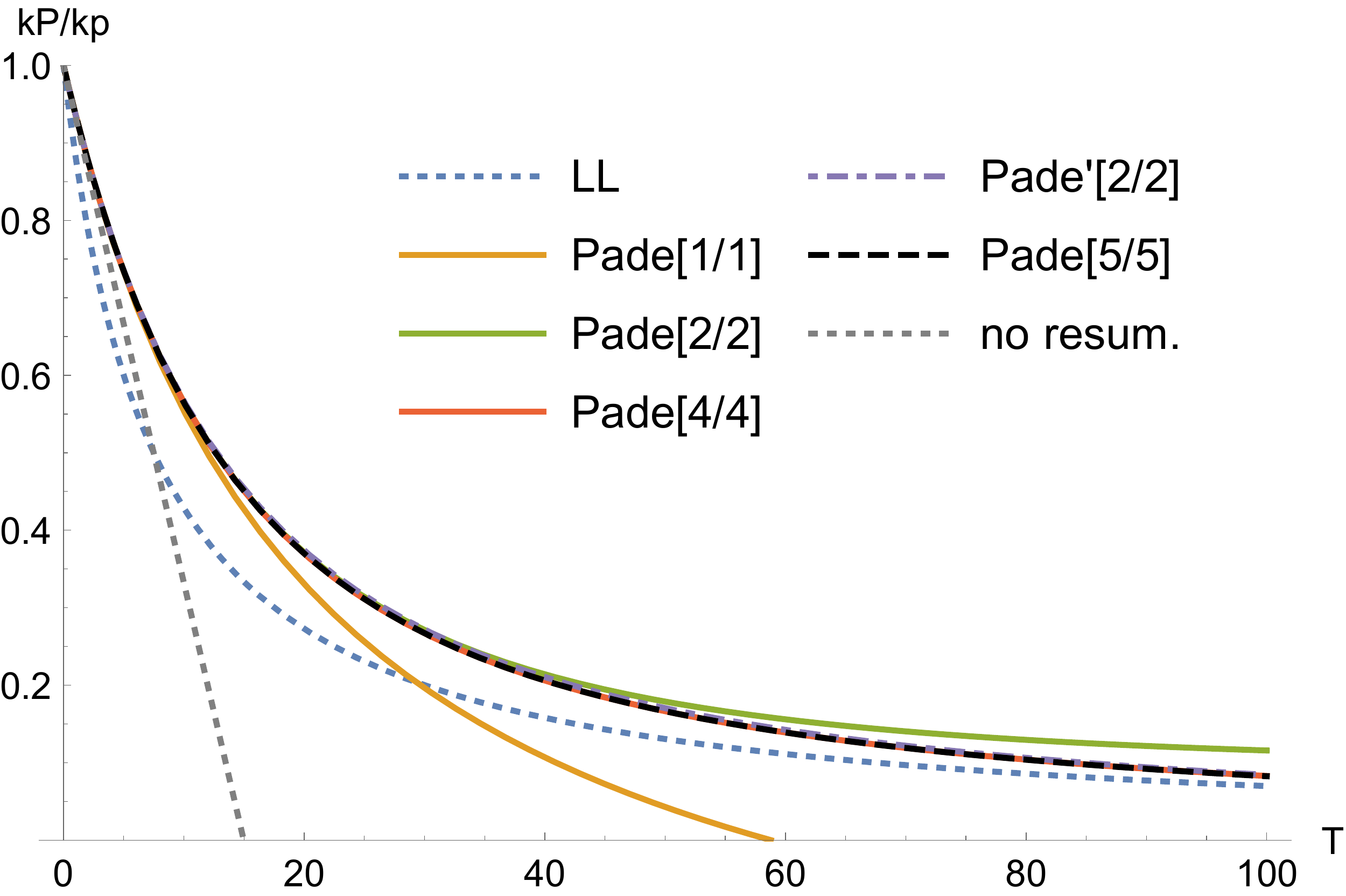}
\includegraphics[width=\linewidth]{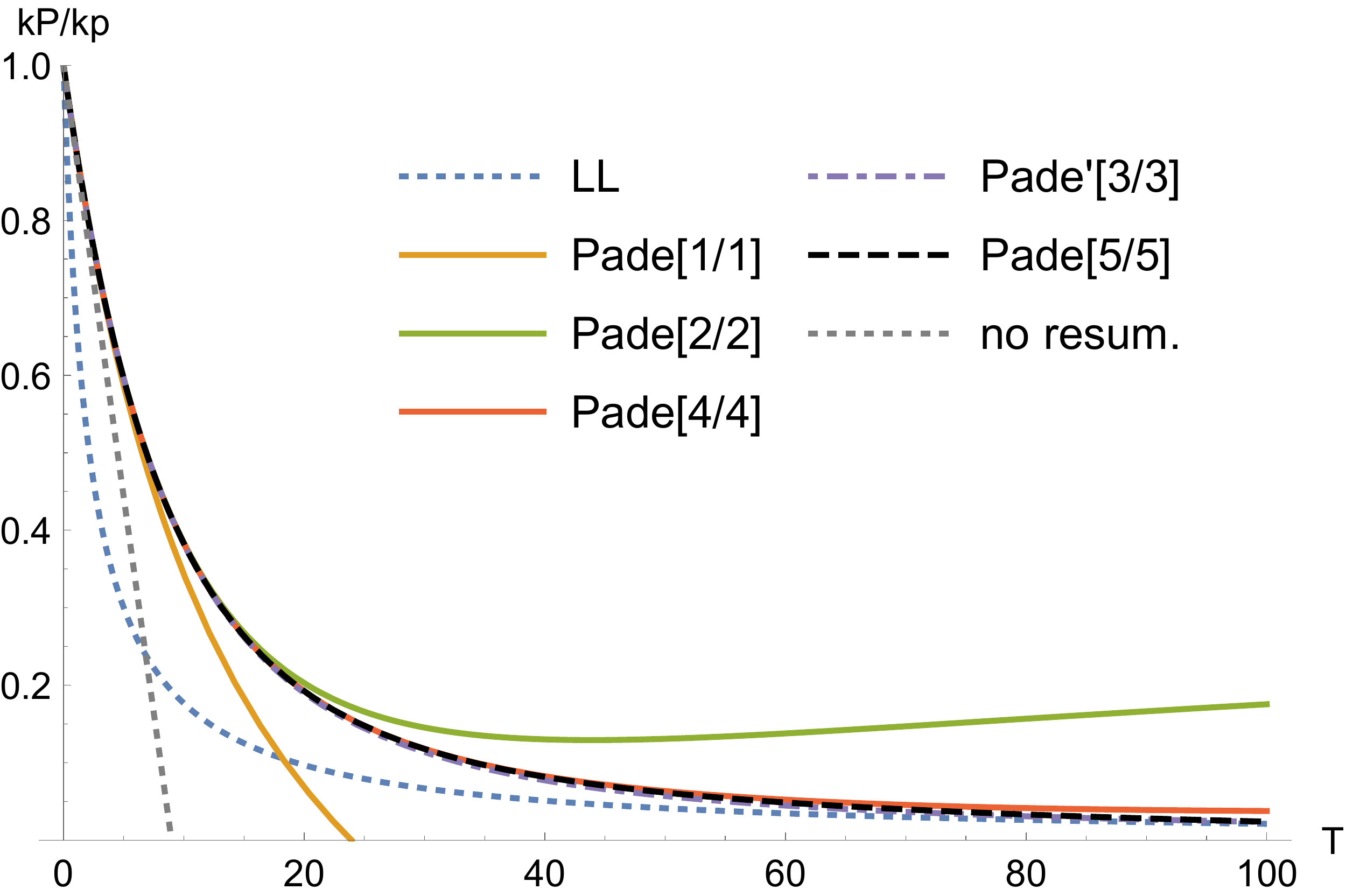}
\caption{Final result for $\chi=0.2$ (upper) and $\chi=0.7$ (lower) as a function of $T=\alpha a_0\Delta\phi$. Pad\'e$[I/I]$ corresponds to the resummation in~\eqref{alphaPadeResum} with $A_i$ and $B_i$ determined from the first $2I$ coefficients $\{1,0\}\cdot{\bf N}^{(n)}$, $1\leq n\leq2I$. Pad\'e$'[I/I]$ is obtained from only $\{1,0\}\cdot{\bf N}^{(n)}$, $1\leq n\leq2I-2$, while the remaining two coefficients are determined from~\eqref{extraAB}. The ``no resum.'' line is just the sum of $\mathcal{O}(\alpha^0)$ and $\mathcal{O}(\alpha)$.}
\label{chi02fig}
\end{figure}

In the second method we expand ${\bf N}^{(1)}(\chi)$ in a power series in $\chi$, which is then used to obtain a power series of ${\bf N}^{(2)}(\chi)$ and so on, see~\ref{AiryIntegrals}. This gives ${\bf N}^{(n)}=\chi^{1+n}\sum_{m=0}^M{\bf a}_m^{(n)}\chi^m$. As illustrated in Fig.~\ref{coefRatioFig}, the coefficients have alternating sign and grow factorially ${\bf a}_m\sim(-1)^mm!$. The $\chi$ expansion is therefore asymptotic with zero radius of convergence and needs to be resummed. 
There is no unique resummation method (unless, of course, one is able to find an exact expression). Different resummations are obtained by matching the series onto different (sums of) special functions. Recent examples involve e.g. the Meijer-G or hypergeometric functions~\cite{Mera:2018qte,Alvarez:2017sza,Torgrimsson:2020mto} (see~\cite{Torgrimsson:2020wlz} for further applications in strong-field QED). Such resummations require fewer terms, but usually some extra information e.g. about the opposite limit (large $\chi$ in our case).
However, in the present case, we can without problems obtain a large number of terms quickly. We can hence use the general Borel-Pad\'e resummation method~\cite{Costin:2019xql,Costin:2020hwg,Caliceti:2007ra,Florio:2019hzn,Dunne:2021acr,Baker1961,BenderOrszag,KleinertPhi4,ZinnJustinBook,Guillou1980}, which only requires the ${\bf a}_m^{(n)}$ coefficients. From the truncated series one first obtains a truncated Borel transform, ${\bf B}_M^{(n)}(t)=\sum_{m=0}^M(1/m!){\bf a}_m^{(n)}t^m$. Next we project with the initial Stokes vector, $a_m^{(n)}:={\bf N}_0\cdot{\bf a}_m^{(n)}$,
and construct a Pad\'e approximant $PB[I/J](t)=\sum_{i=0}^{I} c_i t^i/(1+\sum_{j=1}^J d_j t^j)=B_N(t)+\mathcal{O}(t^{N+1})$. The result is then obtained from the following Laplace integral
\be
{\bf N}_0\cdot{\bf N}_{\rm re}^{(n)}(\chi)=\chi^{1+n}\int_0^\infty\frac{\ud t}{\chi}e^{-t/\chi}PB[I/J](t) \;.
\ee

Using either of these two approaches we obtain a set of functions, ${\bf N}^{(n)}(\chi)$, for $0<\chi<\chi_{\rm max}$. The result for an unpolarized initial state is shown in Fig.~\ref{alphaRatiosFig}. In the low $\chi$ limit we find convergence towards the classical prediction. 



\begin{figure}
\includegraphics[width=\linewidth]{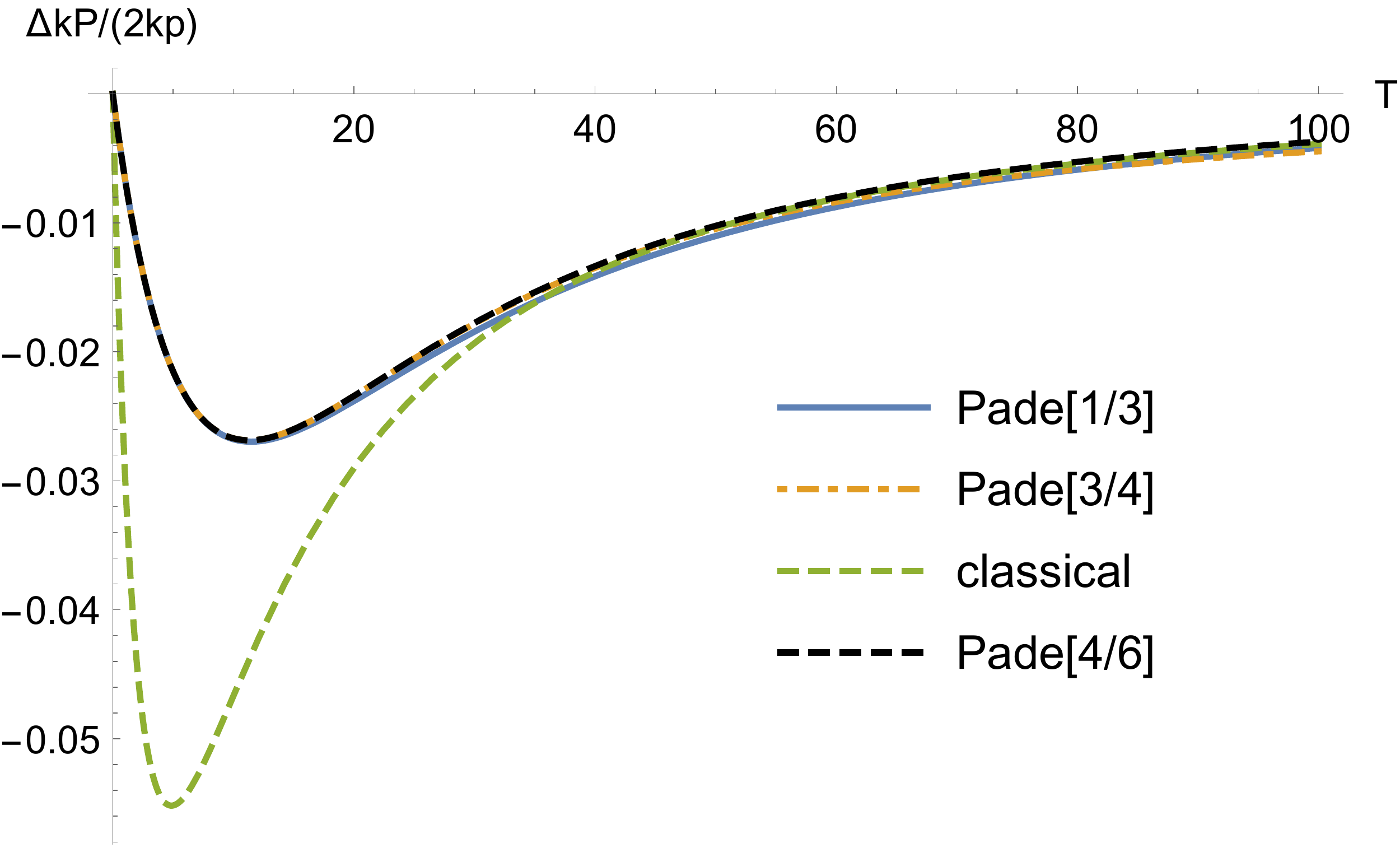}
\caption{Difference in the final momentum for initial state with spin up and down and $\chi=0.2$.}
\label{spinDiffFig}
\end{figure}

In general one would also expect the $\alpha$ expansion to be asymptotic. However, an approximation does not have be asymptotic, see e.g.~\cite{Ritus:1998jm,Affleck:1981bma,Huet:2018ksz,Dunne:2021acr,Karbstein:2019wmj} for the weak and strong field approximations of the QED effective action. (See~\cite{Mironov:2020gbi,Edwards:2020npu} for other recent $\alpha$ resummations.) 
In classical RR, LAD leads to an asymptotic series~\cite{Zhang:2013ria}, while LL has a finite radius of convergence. 
Since the coefficients we have calculated suggest a finite radius of convergence, we propose to resum the $\alpha$ expansion with a Pad\'e approximant
\be\label{alphaPadeResum}
a_0\langle kP\rangle\approx a_0\langle kP\rangle_{\rm re}=\chi+\frac{\sum_{i=1}^IA_i(\chi)T^i}{1+\sum_{j=1}^J B_j(\chi)T^j} \;,
\ee
where $A_i$ and $B_i$ are obtained by matching $a_0\langle kP\rangle_{\rm re}=\chi+\sum_{n=0}^{I+J}T^n{\bf N}_0\cdot{\bf N}^{(n)}+\mathcal{O}(T^{I+J+1})$.
Since we expect a finite limit for $T\gg1$, we take $I=J$. This makes it possible for $\langle kP\rangle_{\rm re}\to0$ as $T\to\infty$, which is what we expect (cf.~\eqref{largeTclassical}). If we do not impose this limit, then we can take $|A_I/(\chi B_I)+1|$ as an upper-limit estimate on the relative error (the error goes to zero as $T\to0$). 
Alternatively, we can demand $\langle kP\rangle_{\rm re}=\mathcal{O}(1/T)$ for $T\gg1$, which fixes $A_I=-\chi B_I$. LL predicts that the leading order in $T\gg1$ is independent of the initial momentum~\eqref{largeTclassical}. If we assume that that holds in general, then the $\mathcal{O}(1/T)$ term must be the same as the classical prediction, which implies
\be\label{extraAB}
A_I=-\chi B_I \qquad
B_I=\frac{2}{3}(A_{I-1}+\chi B_{I-1})  \;.
\ee

Fig.~\ref{chi02fig} shows that the resummation~\eqref{alphaPadeResum} converges quickly. On the scale of this plot, the $I=4$ and $I=5$ resummations are virtually indistinguishable, which are obtained from the first eight and ten ${\bf N}^{(n)}$ terms. For short to moderately large $T$, we see that the classical prediction overestimates the effect of RR, as expected~\cite{Burton:2014wsa}. However, for larger $T$ the classical and quantum predictions seem to converge. This is what one would expect if the leading order at $T\gg1$ is independent on the initial momentum. This motivates us to use the modified Pad\'e approximant based on~\eqref{extraAB}. With the two extra conditions in~\eqref{extraAB} we indeed find an even faster convergence, with a decent approximation already with $I=2$, i.e. using only the $\mathcal{O}(\alpha)$ and $\mathcal{O}(\alpha^2)$ terms. 

These resummations can be compared with the solution to the integro-differential equation corresponding to~\eqref{recursionLCF}, i.e.
\be\label{integroDiffLCF}
\frac{\partial{\bf N}}{\partial T}=\int_0^1\frac{\ud q}{\chi}({\bf M}^{\rm C}\cdot{\bf N}(\chi[1-q])+{\bf M}^{\rm L}\cdot{\bf N}(\chi)) \;,
\ee 
where ${\bf N}=\sum_{n=0}^\infty T^n{\bf N}^{(n)}$ and ${\bf N}(T=0)=\chi\{1,0\}$. We have solved~\eqref{integroDiffLCF} numerically with the Euler method and found good agreement with the resummations above. However, it takes much more time to solve~\eqref{integroDiffLCF} because we need $T_{\rm max}/dT$ steps, which is typically more than the ten (or fewer) steps we needed in the resummation approach. 

At $T\gg1$ we have the ansatz ${\bf N}\approx \{c/T+\mathcal{O}(1/T^2),\mathcal{O}(1/T^2)\}$, so $\partial{\bf N}/\partial T=\mathcal{O}(1/T^2)$, which gives a condition for $c$ since the right-hand side of~\eqref{integroDiffLCF} is not automatically $\mathcal{O}(1/T^2)$. As mentioned, we expect $c$ to be independent of $\chi$, and now we can confirm that this is consistent with~\eqref{integroDiffLCF}. 



In order to obtain ${\bf N}^{(n)}$ from ${\bf N}^{(n-1)}$ we need to calculate both components of ${\bf N}^{(n-1)}$, even if we at the end are only interested in unpolarized particles. Hence, in calculating the results above we have obtained the necessary information to study polarized initial state as a byproduct. 
We begin with the leading order for $\chi\ll1$. With the ansatz ${\bf N}^{(n-1)}\approx\chi^n\{A_{n-1}+C_{n-1}\chi,B_{n-1}\chi\}$ we find ${\bf N}^{(n)}\approx\chi^{n+1}\{-(2/3)A_{n-1}+\mathcal{O}(\chi),-(A_{n-1}+[2/3]B_{n-1}[1+1/n])\chi\}$. Since $A_n=(-2/3)^n$ we find a compact recursive formula for the spin dependence,
\be
{\bf N}^{(n)}_1=-\chi^{2+n}\left[\left(-\frac{2}{3}\right)^{n-1}+\frac{2}{3}\left(1+\frac{1}{n}\right){\bf N}^{(n-1)}_1\right] \;,
\ee
where ${\bf N}^{(n)}_1=\{0,1\}\cdot{\bf N}^{(n)}$. This is an example of a first-order difference equation which can be solved with a general method, see e.g. Eq.~(2.2.7) in~\cite{BenderOrszag}. 
We find
\be
{\bf N}^{(n)}_1=-\left(-\frac{2}{3}\right)^{n-1}\chi^{2+n}(1+n)(H_n-1) \;,
\ee
where $H_n$ is the $n$-th harmonic number. This compact formula allows us to find an exact resummation,
\be\label{spinDiffAll}
\frac{a_0}{2}k\Delta\langle P\rangle=\sum_{n=1}^\infty{\bf N}^{(n)}_1T^n=-\frac{3}{2}\chi^2\frac{\ln\left[1+\frac{2}{3}\chi T\right]}{\left(1+\frac{2}{3}\chi T\right)^2} \;,
\ee
where $\Delta\langle P\rangle=\langle P\rangle({\bf N}_0=\{1,1\})-\langle P\rangle({\bf N}_0=\{1,-1\})$.
This all-order result has the same radius of convergence as the solution to LL. However, for $\Delta\langle P\rangle$ we see that the convergence-limiting singularity is a combination of a pole and a branch point. We also see that $\Delta\langle P\rangle$ too becomes independent of $\chi$ to leading order in $T\gg1$, although this time the next-to-leading order is only logarithmically suppressed. Another difference is that~\eqref{spinDiffAll} is non-monotonic with a maximum $|\Delta\langle P\rangle|$ at $T\sim1/\chi$.

The full quantum result can be obtained as described above. Although~\eqref{spinDiffAll} might suggest using a resummation involving logarithms, we still find a fast convergence with a Pad\'e approximants as in~\eqref{alphaPadeResum}, except that $P_\LCm>0$ implies that $|\Delta\langle P\rangle|$ must be smaller than $\langle P\rangle_\uparrow+\langle P\rangle_\downarrow$ and for that to hold at large $T$ we need $J>I$. In practice, different choices of $I$ and $J$ can give good approximations, and a ``wrong'' choice just means that we need to include more terms.
Fig.~\ref{spinDiffFig} shows that the classical prediction overestimates the peak of $|\Delta\langle P\rangle|$ by about a factor of two for $\chi=0.2$, but is close to the quantum result for large $T$.


In conclusion, we have derived a recursion formula for calculating the expectation value of an electron in a plane wave background field, where $\mathcal{O}(\alpha^n)$ is obtained by multiplying the spin dependent $\mathcal{O}(\alpha^{n-1})$ result with a Mueller matrix for nonlinear Compton scattering and the corresponding loop and then integrating over the photon momentum. In the classical limit we find the solution to LL to all orders. We have shown that quantum RR can be obtained either by constructing interpolation functions of each order, or by expanding each $\mathcal{O}(\alpha^n)$ in $\chi$ and then resumming these asymptotic series with e.g. the Borel-Pad\'e method. We have also found that the $\alpha$ expansion obtained in this way can be be resummed with Pad\'e approximants. The fast convergence of these approximants is encouraging for the generalization to non-constant fields. Our approach takes the spin into account using Mueller matrices, so we have also studied spin dependent RR.

\appendix

\section{Classical limit of remaining components}\label{ClassicalLimitPerpPlus}

In the main text we showed that the classical limit of $\langle P_\LCm\rangle$ agrees with the solution to LL. Now we will extend this to the transverse components. Since this introduces more variables, we use for convenience a more compact notation. Let $\langle{\sf p}'-\hat{\sf a}\rangle_C$ and $\langle{\sf p}-\hat{\sf a}\rangle_L$ be the contribution to $\langle{\sf P}\rangle^{(1)}$ coming from Compton scattering and the loop, respectively, where ${\sf p}'={\sf p}-{\sf l}$ is the momentum of the electron after emitting a photon with momentum ${\sf l}$, and $\hat{\sf a}={\sf a}(\hat{\sigma})$ where $\hat{\sigma}$ is some upper limit for $\sigma$.  
In general we would need to use the Mueller matrices for $\langle P_{\LCperp}\rangle^{(1)}$. However, to compare with classical physics we only have to consider $M=\{1,{\bf0}\}\cdot{\bf M}\cdot\{1,{\bf0}\}$. We have
\be
\langle{\sf P}\rangle^{(1)}({\sf p},-\infty)=\langle{\sf p}'-\hat{\sf a}\rangle_C+\langle{\sf p}-\hat{\sf a}\rangle_L=\langle-{\sf l}\rangle_C \;,
\ee
which corresponds to Eq.~(9) in~\cite{Ilderton:2013tb}.
To obtain higher orders we let the lower integration limit for the $\sigma'$ integral be $\sigma$ rather than $-\infty$. We can obtain this from~\cite{Ilderton:2013tb,Ilderton:2013dba}
\be
\langle{\sf P}\rangle^{(1)}({\sf p},\sigma)=\Delta \pi'_\LCperp(\hat{\sigma})-\Delta\int_\sigma^{\hat{\sigma}}\ud\sigma'{\bf a}'^2(p-a)_\LCperp \;.
\ee
We could drop the $\pi'$ term since it does not contribute to the leading order for long pulse length or high intensity, and we can only expect the Mueller matrix/incoherent product approach to give the leading order. However, we will keep it since at least in this case doing so allows us to obtain all terms in the solution to LL.
To obtain $\mathcal{O}(\alpha^2)$ we should prepend a Compton scattering or a loop step before the first-order result, 
\be
\langle{\sf P}\rangle^{(2)}({\sf p},\sigma)=\langle\langle{\sf P}\rangle^{(1)}({\sf p}',\sigma')-\langle{\sf P}\rangle^{(1)}({\sf p},\sigma')\rangle_C \;,
\ee
where the $\sigma'$ integral is restricted by $\sigma'>\sigma$. Since $\pi'_\LCperp=-a'_\LCperp$ is independent of the momentum, the first term, $\Delta \pi'_\LCperp(\hat{\sigma})$, depends only on the longitudinal momentum and with a linear scaling, and can therefore be treated in exactly the same way as the above calculation for $\langle kP\rangle$. For the second term we have by expanding to linear order in the photon momentum
\be
kp'(p'-a)_\LCperp-kp(p-a)_\LCperp\approx-kl(p-a)_\LCperp+kp(-l)_\LCperp \;,
\ee 
so we only need to calculate $\langle-{\sf l}\rangle_C$ which we have already obtained for $\langle{\sf P}\rangle^{(1)}$, we just have to remember the $\sigma$ ordering. In general we find
\be\label{PclassRec}
\langle{\sf P}\rangle^{(n)}({\sf p},\sigma)\approx-\int_\sigma^\infty\ud\sigma'\frac{\partial\langle{\sf P}_a\rangle^{(1)}}{\partial\sigma'}\frac{\partial\langle{\sf P}\rangle^{(n-1)}}{\partial {\sf p}_a} \;,
\ee
where a sum over $a=-,1,2$ is implied.
For the contribution to $\langle P_\LCperp\rangle^{(n)}$ from the part without $\pi'$ we find
\be
\begin{split}
&(n-1)!\Delta^n\int_\sigma\ud\sigma_1\int_{\sigma_1}\ud\sigma_2...\int_{\sigma_{n-1}}\ud\sigma_n \\
&\times[I_\LCperp(\sigma_1)I(\sigma_2)...I(\sigma_n)+I(\sigma_1)I_\LCperp(\sigma_2)...I(\sigma_n) \\
&\hspace{1cm}+...+I(\sigma_1)I(\sigma_2)...I(\sigma_n)_\LCperp] \\
&=\left(\Delta\int_\sigma I\right)^{n-1}\Delta\int_\sigma I_\LCperp \;,
\end{split}
\ee
where $I=-{\bf a}'^2$ and $I_\LCperp=-{\bf a}'^2(p-a)_\LCperp$.
Thus, we again find a geometric series and by resumming this we find~\eqref{mperpLL}.

The classical limit of $\langle P_\mu\rangle$ is on shell, i.e. $\langle P_\LCp\rangle$ can be obtained from~\eqref{mperpLL}. So, the following, direct calculation of $\langle P_\LCp\rangle$ can be seen as an additional check. We now drop the $\pi'$ term. 
We will use~\eqref{PclassRec} (where the sum is still over $a=-,1,2$).
From
\be
\langle P_\LCp\rangle^{(1)}=-\Delta\int{\bf a}'^2\left(\pi_\LCp-\frac{\hat{\pi}\pi}{kp}k_\LCp\right) \;,
\ee
where $\hat{\pi}\pi=1+(1/2)(\hat{\bf a}-{\bf a})^2$ is independent of $p_\mu$, we find $\partial \langle P_\LCp\rangle^{(1)}/\partial p_\LCm=0$, so only $a=1,2$ contribute and
\be
\langle P_\LCp\rangle^{(2)}\approx\frac{\Delta^2}{2p_\LCm}\int\ud\sigma_1\int_{\sigma_1}\ud\sigma_2 I_\LCperp(\sigma_1)I_\LCperp(\sigma_2) \;.
\ee
This is quite different from $\langle P_\LCp\rangle^{(1)}$, but at third order we find
\be
\begin{split}
\langle P_\LCp\rangle^{(3)}\approx&\frac{\Delta^3}{2p_\LCm}\int\ud\sigma_1\int_{\sigma_1}\ud\sigma_2\int_{\sigma_2}\ud\sigma_3[I(\sigma_1)I_\LCperp(\sigma_2)I_\LCperp(\sigma_3) \\
&+I_\LCperp(\sigma_1)I(\sigma_2)I_\LCperp(\sigma_3)+I_\LCperp(\sigma_1)I_\LCperp(\sigma_2)I(\sigma_3)] \;,
\end{split}
\ee
from which we start to see a pattern. Thus, after $\langle P_\LCp\rangle^{(2)}$ we find
\be
\begin{split}
\langle P_\LCp\rangle^{(n)}&\approx\frac{(n-2)!\Delta^n}{2p_\LCm}\int\ud\sigma_1...\int_{\sigma_{n-1}}\ud\sigma_n\sum I(\sigma_1)... \\
\times&I(\sigma_{i-1})I_\LCperp(\sigma_1)I(\sigma_{i+1})...I(\sigma_{j-1})I_\LCperp(\sigma_j)I(\sigma_{j+1})... \;,
\end{split}
\ee  
where the sum is over all pairs $\{\sigma_i,\sigma_j\}$, which can be rewritten as
\be
\langle P_\LCp\rangle^{(n)}\approx\left(\Delta\int\ud\sigma\, I\right)^{n-2}\frac{\Delta^2}{4p_\LCm}\left(\int\ud\sigma\, I_\LCperp\right)^{2} \;.
\ee
Thus, we again find a geometric series and by resumming it we find
\be
\langle P_\LCp\rangle\approx\hat{\pi}_\LCp-\Delta\int{\bf a}'^2\left[\pi_\LCp-\frac{\hat{\pi}\pi}{2p_\LCm}\right]+\frac{\Delta^2}{4p_\LCm}\frac{\left[\int\ud\sigma\, {\bf a}'^2\pi_\LCperp\right]^{2}}{1+\Delta\int\sigma{\bf a}'^2} \;,
\ee
which is just what is needed for $\langle P_\mu\rangle$ to be on shell.

\section{$\chi$ expansion}\label{AiryIntegrals}

The $\chi$ expansion can be obtained by changing integration variables from $q=\chi\gamma/(1+\chi\gamma)$ to $\gamma$, which removes $\chi$ from the argument of the Airy functions, and then simply expanding the integrand. The resulting integrals are
\be
\begin{split}
\int_0^\infty\ud\gamma\, \gamma^n\frac{\text{Ai}(\gamma^{2/3})}{\gamma^{1/3}}&=
\frac{3^{\frac{1}{2}+n}}{4\pi}\Gamma\left[\frac{1}{3}+\frac{n}{2}\right]\Gamma\left[\frac{2}{3}+\frac{n}{2}\right] \\
&=\left\{\frac{\sqrt{3}}{4},1,\frac{35\sqrt{3}}{16},20,\dots\right\}
\end{split}
\ee
\be
\begin{split}
\int_0^\gamma\ud\gamma\, \gamma^n\frac{\text{Ai}'(\gamma^{2/3})}{\gamma^{2/3}}&=-\frac{3^{\frac{1}{2}+n}}{4\pi}\Gamma\left[\frac{1}{6}+\frac{n}{2}\right]\Gamma\left[\frac{5}{6}+\frac{n}{2}\right] \\
&=\left\{-\frac{1}{2},-\frac{5\sqrt{3}}{8},-4,-\frac{385\sqrt{3}}{32},\dots\right\}
\end{split}
\ee
\be
\begin{split}
\int_0^\gamma\ud\gamma\, \gamma^n\text{Ai}_1(\gamma^{2/3})&=\frac{3^{\frac{1}{2}+n}\Gamma\left[\frac{5}{6}+\frac{n}{2}\right]\Gamma\left[\frac{7}{6}+\frac{n}{2}\right]}{2\pi(1+n)} \\
&=\left\{\frac{1}{3},\frac{35}{24\sqrt{3}},\frac{10}{3},\frac{1001}{32\sqrt{3}},\dots\right\}
\end{split}
\ee
where
\be
\text{Ai}_1(\xi)=\int_\xi^\infty\ud x\,\text{Ai}(x) \;.
\ee
These gamma functions are responsible for the factorial growth of the coefficients in the $\chi$ expansion.

Note that the terms in the $\chi$ expansion are completely analytical, but higher orders involve integers with a very large number of digits. Hence, for finding Pad\'e approximants of the Borel transform and for performing the Laplace integral, one may have to work with very high precision.

\end{document}